\documentclass[11pt]{article}
\usepackage[dvips]{graphicx}
\usepackage{enumerate}
\usepackage{float}
\usepackage{latexsym,amsmath,mathrsfs,bm,amssymb,color,hiroshima}
\newtheorem{theorem}{Theorem}[section]

\newtheorem{lemma}[theorem]{Lemma}
\newtheorem{corollary}[theorem]{Corollary}
\newtheorem{definition}[theorem]{Definition}
\newtheorem{example}[theorem]{Example}
\newtheorem{remark}[theorem]{Remark}

\newcommand{\proof}{Proof:\ }
\newcommand{\qed}{\hfill $\Box$ \par\medskip}
\newcommand{\s}{{\rm Spec}}
\renewcommand{\ba}{\begin{align}}
\newcommand{\ea}{\end{align}}
\renewcommand{\LE}{L^2_0(\RR)}
\newcommand{\LO}{L^2_1(\RR)}
\newcommand{\bd}[1]{\begin{definition}\label{#1}}
\newcommand{\ed}{\end{definition}}

\newcommand{\LK}{\kL}
\newcommand{\TG}{T_{G}}
\newcommand{\TGG}{T_{\rm POVM}}

\renewcommand{\lr}[2]{\langle #1| #2 \rangle}

\newcommand{\AAA}{\log\frac{1+\alpha}{1-\alpha}}
\newcommand{\BBB}{\log\frac{1+\beta}{1-\beta}}
\newcommand{\xz}{\xi_{z}}
\newcommand{\xa}{\xi_{i\alpha }}
\newcommand{\xae}{\xi_{i\alpha ,\,\eps}}

\newcommand{\xbe}{\xi_{i,\,\eps}}

\newcommand{\xce}{\xi_{i\beta,\,\eps}}
\newcommand{\tax}{D_{i\alpha }}
\newcommand{\taxe}{D_{i\alpha ,\,\eps}}
\newcommand{\tbxe}{D_{i\beta,\,\eps}}
\newcommand{\taxo}{D_{i\alpha ,\,1}}
\newcommand{\tbxo}{D_{i\beta,\,1}}
\newcommand{\tzx}{D_{z}}

\makeatletter
\@addtoreset{equation}{section}
\makeatother

\begin{document}
\title{\sc Time Operators of Harmonic Oscillators and Their Representations\footnote{We dedicate this paper in honor of Professor Asao Arai on his 70th birthday.}}
\author{Fumio Hiroshima \\
Faculty of Mathematics, Kyushu University\footnote{hiroshima@math.kyushu-u.ac.jp} \\ 
and \\ 
Noriaki Teranishi\\ 
Faculty of Science, 
Hokkaido University\footnote{teranishi@math.sci.hokudai.ac.jp}}
\date{\today}
\maketitle

\begin{abstract}
A time operator $\hat T_\eps$ of the one-dimensional harmonic oscillator 
$ \hat h_\eps=\half(p^2+\eps q^2)$ is rigorously constructed. 
It is formally expressed as 
$ \hat T_\eps=\half\frac{1}{\sqrt \eps } (\arctan (\sqrt \eps \hat t_0)+\arctan (\sqrt \eps \hat t_1))$ with $\hat t_0=p^{-1}q$ and $\hat t_1=qp^{-1}$. 
It is shown that the canonical commutation relation 
$[h_\eps, \hat T_\eps ]=-i\one$ holds true on a dense domain 
in the sense of sesqui-linear forms, and the limit of $\hat T_\eps $ as $\eps\to 0$ is shown. 
Finally a matrix representation of $\hat T_\eps$ and its analytic continuation are given. 

\end{abstract}


\section{Introduction}
\subsection{Time operators}
\label{TT}
In this paper, we are concerned with time operators of 
the one-dimensional harmonic oscillator. 
Time operators are defined in this paper 
as conjugate operators for self-adjoint operators. 
We begin with defining time operators. 
Let $[A,B]$ be the commutator of linear operators $A$ and $B$ defined by 
$$[A,B] =AB-BA.$$
\begin{definition} Let $A$ be self-adjoint and $B$ symmetric on a Hilbert space $\cH$. 
If $A$ and $B$
satisfy 
the canonical commutation relation:
\begin{align*}
[A,B]=-i\one\end{align*}
on a non-zero subspace $D_{A,B}\subset \rD(AB)\cap \rD(BA)$, 
then $B$ is called a {time operator} of $A$. 
\end{definition}
Although the term \lq\lq time operator" may not be appropriate at some situations, we follow the conventions and use this terminology throughout this paper.
We show several examples of time operators. 

{(Position-Momentum)}
Let $p=\frac{1}{i}\frac{d}{dx}$ and $q=M_x$ be the multiplication by $x$. 
Both $p$ and $q$ are self-adjoint operators on 
\begin{align*}
&\rD(p)=\left\{f\in\LR\ \middle|\ k\hat f(k)\in \LR\right\},\\
&\rD(q)=\left\{f\in\LR\ \middle|\ xf\in \LR\right\},
\end{align*}
respectively, where $\hat f$ denotes the 
Fourier transform of $f\in \LR$, 
which is given by 
$\hat f(k)=\frac{1}{\sqrt{2\pi}}\int _\RR f(x)e^{-ikx}d x$. 
Hence the canonical commutation relation 
\begin{align}
\label{pq}
[p,q]=-i\one
\end{align}
holds on $\rD(pq)\cap \rD(qp)$. 
Thus 
$q$ is a time operator of $p$, 
and 
$\rD(pq)\cap \rD(qp)$ is dense in $\LR$.

{(Energy-Time: continuous case)}
The operator $\half p^2$ describes the energy of the one-dimensional free quantum particle with the unit mass and the spectrum of $\half p^2$ is purely absolutely continuous, i.e., 
$\s(\half p^2)=[0,\infty)$. 
Let 
\begin{align}\label{time}
\hat T_{\rm AB}=\half \left(p^{-1}q+qp^{-1}\right).
\end{align}
$\hat T_{\rm AB}$ is called the {Aharonov-Bohm operator} \cite{AB61} or the time of arrival operator. 
It holds that 
\begin{align}\label{times2}
\left[\half p^2,\hat T_{\rm AB}\right]=-i\one
\end{align}
on a dense domain. 
Then $\hat T_{\rm AB}$ is a time operator of $\half p^2$. 

{(Energy-Time: discrete case)}
Let $\half (p^2+q^2)$ be the one-dimensional harmonic oscillator which is the main object in this paper. 
The spectrum of $\half (p^2+q^2)$ is purely discrete, i.e., 
$\s\lk \half (p^2+q^2)\rk=\{n+\half\}_{n=0}^\infty$. 
Let
\begin{align}
\label{hatt}
\hat T=\half\left(\arctan (p\f q)+\arctan (qp\f)\right). 
\end{align}
$\hat T$ is called the {angle operator}. 
The name comes from the fact below; 
in the classical phase space the angle $\theta$ between $(p_c,q_c)$ and $p_c$-axis is described by $\theta=\arctan(q_c/p_c)$. 
It can be seen 
that 
$\hat T$ and $\half(p^2+q^2)$ {\it formally} satisfy 
\begin{align}
\label{ht}
\left[\half (p^2+q^2),\hat T\right]=-i\one
\end{align}
on some domain. Then $\hat T$ is a time operator of $\half (p^2+q^2)$. However, as far as we know, there are no rigorous definitions of 
both $\arctan(p\f q)$ and $\arctan(qp\f)$. 

{(Number-Phase)}
Let $a=q+ip$ and $\add=q-ip$. 
Then $N=\add a$ is called the number operator. 
It holds that $[a,\add]=\one$. 
In physics literatures a symmetric 
operator $\hat\phi$ satisfying 
\begin{align}
 \label{np}
 \left[N,\hat\phi\right]=-i\one
\end{align}
is called the {phase operator}, 
and {\it formally} it is described as 
\begin{align}
\label{phase}
\hat\phi=\frac{i}{2}(\log a-\log\add).
\end{align}
Thus 
$\hat\phi$ is a time operator of $N$. 
However, there are also no rigorous definitions of 
both $\log a$ and $\log\add$.

Four canonical commutation relations \eqref{pq}, \eqref{times2}, \eqref{ht} and \eqref{np} 
have been studied historically in theoretical physics so far, 
but 
\eqref{ht} and \eqref{np} are crucial to investigate as 
operator equalities from a mathematical point of view. 
In this paper we study \eqref{ht}, 
and 
in the second paper \cite{HT22} we study \eqref{np} and 
relationships between \eqref{hatt} and \eqref{phase}. 

In this paper we discuss two time operators of the one-dimensional harmonic oscillator. 
One is the { angle operator} and the other 
an operator defined by a { positive operator valued measure} (POVM).
The purpose of this paper is to consider properties of 
both the angle operator and the operator defined by POVM, rigorously, 
and to show relationships between 
time operators 
of Hamiltonians with purely discrete spectrum 
 and 
 time operators 
of Hamiltonians with purely absolutely continuous spectrum under the firm mathematical frame. 
The later can be reduced to consider a limit of the family of 
angle operators $T_\eps$ defined below as $\eps\to0$.

\subsection{Aharonov-Bohm operator, angle operators and POVM}

Within the long-running discussion on time in quantum mechanics, 
Aharonov and Bohm introduced the operator $\hat T_{\rm AB}$ \cite{AB61}. 
They considered the quantization of functions of the classical momentum $P$ and the classical position $Q$, 
which provides us with the time of arrival at a point $y$ of a particle that at the instant $t=0$ has position $x$ and momentum $P$, i.e., 
\begin{align}
\label{arrival}
\mbox{time of arrival} =\frac{x-y}{P}.
\end{align}
The classical expression \eqref{arrival} of the time of arrival 
suggests to define the symmetric time operator of the free Hamiltonian $\half p^2$ 
in quantum mechanics by
$\hat T_{\rm AB}$. 
It is established that $\hat T_{\rm AB}$ is symmetric but 
no self-adjoint extensions exist, and 
it satisfies 
 $$\left[\half p^2, \hat T_{\rm AB}\right]=-i\one$$
on a dense domain. 
See e.g., \cite{miy01} and \cite[Theorems 4.21 and 4.22]{ara20}.

In this paper we are concerned with time operators 
of the one-dimensional harmonic oscillator with parameter $\eps$: 
$$\half (p^2+\eps q^2),\quad 0<\eps\leq1.$$
Operator $\half (p^2+\eps q^2)$ is self-adjoint on $\rD(p^2)\cap \rD(q^2)$. 
We shall construct $\hat T_\eps$ such that 
$$\left[\half (p^2+\eps q^2),\hat T_\eps\right]=-i\one.$$ 
By a heuristic argument we see in Section \ref{heuristic}, 
formally $\hat T_\eps$ is given by 
$$\hat T_\eps=\half \frac{1}{\sqrt \eps}
\left(\arctan (\sqrt\eps p^{-1} q)+
\arctan (\sqrt\eps q p^{-1})\right).$$

One standard way of constructing time operators is to apply 
POVMs. The details are demonstrated in Section \ref{povm2}. 
Let $P$ be a POVM 
on a measurable space $(\Omega,\cB)$ 
associated with a self-adjoint operator $H$. 
We define 
\[\TGG =\int_{\Omega} t dP_t.\] 
Then $\TGG $ is a time operator of $H$. 
By using the POVM associated with $\half(p^2+q^2)$ 
one can also construct a time operator $\TG$ of $\half(p^2+q^2)$. 
See \eqref{p1} and \eqref{p2}. 

\subsection{Momentum representation}
In what follows we 
take the momentum representation instead of the position representation. 
Let $F\colon L^2(\RR_x)\to L^2(\RR_k)$ be the Fourier transformation. 
Then \[Fp F\f=M_k,\quad Fq F\f=+i\frac{d}{dk},\]
where $M_k$ is the multiplication by $k$ on $L^2(\mathbb R_k)$:
\begin{align*}
 &\rD(M_k)=\left\{f\in L^2(\mathbb R_k)\ \middle|\ kf(k)\in L^2(\mathbb R_k)\right\},\\
 &M_kf(k)=kf(k).
\end{align*} 
Since we work in $L^2(\RR_k)$ for $L^2(\RR_x)$ 
in what follows because of taking the momentum representation, 
instead of notations $L^2(\RR_k)$, $M_k$ and $i\frac{d}{dk}$, 
we rewrite them as 
$L^2(\RR), q$ and $-p$, respectively, unless confusions may arise. 
Thus $[p,q]=-i\one$ also holds true in the momentum representation. 
Let us recall the Aharonov-Bohm operator in the momentum representation,
which is a time operator of $\half q^2$. 
Since $q$ is an injective self-adjoint operator, the inverse operator $q^{-1}$ 
exists as a self-adjoint operator. 
In particular, the domain of $q^{-1}$ is dense. 
Let 
\begin{align}
t_0=q^{-1}p
\end{align}
with the domain
$\rD(t_0)=\{f\in \rD(p)\mid pf\in \rD(q^{-1})\}$. 
The operator $t_0$ is densely defined but not symmetric. 
We also define $t_1$ by 
\begin{align}
t_1=pq^{-1}
\end{align}
with the domain
$\rD(t_1)=\{f\in \rD(q)\mid qf\in \rD(p^{-1})\}$. 
As with $t_0$, $t_1$ is also densely defined but not symmetric. 
In the momentum representation, 
the Aharonov-Bohm operator is given by 
\[T_{\rm AB}=-\half (t_0+t_1).\]
The operator $T_{\rm AB}$ is a densely defined symmetric operator, and it follows that 
\[\left[\half q^2,T_{\rm AB}\right]=-i\one\]
on a dense domain. 

\subsection{Heuristic derivation of angle operators}
In the momentum representation, 
$\hat h_\eps$ is then transformed to 
$$h_\eps=\half(\eps p^2+q^2).$$ 
We shall construct an operator $T_\eps$ such that
$$
[h_\eps,T_\eps]=-i\one.$$
Heuristically we shall derive $T_\eps$. 
We have
$[h_\eps, t_0]=i(\one+\eps t_0^2)$ and then 
it follows that 
$[h_\eps, f(t_0)]=i(\one+\eps t_0^2)f'(t_0)$ for a function $f$ on $\RR$. 
From this we may expect that 
$f'(t_0)=(\one+\eps t_0^2)^{-1}$ and then 
\[f(t_0)=-\frac{1}{\sqrt \eps} \arctan (\sqrt \eps t_0).\]
Symmetrizing $f$, we see that 
\begin{align}
\label{T}
T_\eps=-\frac{1}{2\sqrt \eps}(\arctan (\sqrt \eps t_0)+\arctan (\sqrt \eps t_1)) 
\end{align}
 may be a time operator of $h_\eps$. 
 
Operator $T_\eps$ appears in e.g., \cite{bau83,BG12, CN68,lef69,LLH96,PG89,SG64}. 
In a literature, however it is stated that $\arctan t$ is bounded because of 
$|\arctan x|\leq \pi/2$ for $x\in\RR$. 
 It is however incorrect. 
Since $t$ is unbounded and not symmetric, 
one can not apply the functional calculus for self-adjoint operators 
to define both $\arctan (\sqrt \eps t_0) $ and $\arctan(\sqrt \eps t_1)$.
Then it is not trivial to define 
$\arctan (\sqrt \eps t_\#)$ for $\#=0,1$. 
In this paper we define them by the Taylor series:
\begin{align}
\label{tay}
\arctan (\sqrt \eps t_\#)=\sum_{n=0}^\infty \frac{(-1)^{n}}{2n+1}
(\sqrt \eps t_\#)^{2n+1}.
\end{align}
It is emphasised that $\arctan x =
\sum_{n=0}^\infty (-1)^nx^{2n+1}/(2n+1)$ is valid for $|x|<1$.
Moreover it is not straightforward to specify a {\it dense} domain $\rD$ such that
\[\rD\subset \bigcap_{n=0} ^\infty \left(\rD(t_0^n)\cap \rD(t_1^n)\right)\] and 
for $f\in \rD$, 
$$\sum_{n=0}^\infty \frac{(-1)^n}{2n+1}(\sqrt \eps t_\#)^{2n+1}f\in \LR.$$

\section{Main results}

\subsection{Hierarchy of time operators}
\label{heuristic}
The following hierarchy of classes of time operators is introduced in \cite{AH17}: 
$$\{\mbox {Ultra-strong time}\}\subset 
\{\mbox{Strong time}\}\subset 
\{\mbox{Time}\}\subset
\{\mbox{Weak time}\} \subset 
\{\mbox{Ultra-weak time}\}. 
$$
We explain this hierarchy below. 
Let $A$ and $B$ be self-adjoint. 
If the Weyl relation 
\begin{align}\label{ab}
e^{-is B}e^{-itA}=e^{its}e^{-itA}e^{-is B},\quad s,t\in\RR
\end{align}
holds true, 
then $[A,B]=-i\one$ also holds true, which is deduced by taking derivatives $d/ds$ and $d/dt$ on both sides of \eqref{ab}. Then $B$ is called 
the {ultra-strong time operator} of $A$. 
It is known that the position operator $q$ and the momentum operator $p$ satisfy the Weyl relation.
Furthermore by the von Neumann uniqueness theorem \cite{N31a}, 
if the Weyl relation \eqref{ab} holds, and both $e^{-itA}$ and $e^{-isB}$ are irreducible, then 
$A$ (resp.$B$) is unitarily equivalent to $p$ (resp.~$q$).

A pair $\{A, B\}$ consisting of a self-adjoint operator $A$ and a 
symmetric operator $B$ 
is called a {weak Weyl representation} 
if 
$e^{-itA}\rD(B)\subset\rD(B)$ for all $t\in \RR$ and 
\begin{equation}
B e^{-itA}\psi =e^{-itA} (B+t)\psi,\quad t\in \RR \label{wwr}
\end{equation}
holds true for all $\psi\in \rD(B)$. 
If $ \{A,B\}$ is a weak Weyl representation, 
then 
the canonical commutation relation $[A,B]=-i\one$ also holds true. 
Then $B$ is called the {strong time operator} of $A$. 
It is shown in \cite{miy01} 
that $T_{\rm AB}$ is a strong time operator of $\half q^2$. 
The crucial fact is that 
if $ \{A,B\}$ is a weak Weyl representation, then 
the spectrum of $A$ is purely absolutely continuous. 
In particular there are no eigenvalues of $A$. 
We refer to see \cite{miy01,ara05,ara20,P67} for the comprehensive study 
of weak Weyl representations.

One difficulty of the investigation of time operators 
is to specify their domains. 
Suppose that $[A,B]=-i\one$. 
Let $Af=Ef$ with $E\in\RR$, i.e., $f$ is an eigenvector of $A$. 
Then at least either $f\not\in \rD(B)$ or $Bf\not\in \rD(A)$ holds true. 
To study time operators of self-adjoint operators possessing eigenvalues it is useful to introduce weak time operators and ultra-weak time operators to avoid the domain argument mentioned just above. 

A symmetric operator $B$ is called 
the {weak time operator} of 
a self-adjoint operator $A$ if there exists a non-zero subspace $D\subset \rD(B)\cap \rD(A)$ on which the weak canonical commutation relation holds:
\begin{equation}
(A\phi, B\psi)- (B\phi, A\psi)=-i (\phi, \psi),\quad \phi,\psi\in D. 
\label{w-CCR}
\end{equation}

We can furthermore generalize weak time operators, which is the 
so-called ultra-weak time operator. 
Ultra-weak time operators are the key tool in this paper. 

\begin{definition}[ultra-weak time operator \cite{AH17}]\label{def-uwt}
Let $H$ be a self-adjoint operator on $\cH $ and 
$D_1$ and $D_2$ be non-zero subspaces of $\cH $.
A sesqui-linear form 
\[ \cT \colon D_1 \times D_2\to\CC, \quad 
D_1\times D_2\ni (\phi,\psi)\mapsto \cT [\phi,\psi]\in\CC\]
with the domain $\rD(\cT )=D_1\times D_2$ 
is called { an ultra-weak time operator} of $H$ if
there exist non-zero subspaces $D$ and 
$E$ of $D_1\cap D_2$ such that
the following {\rm (1)--(3)} hold:
\begin{list}{}{}
\item[\rm (1)] $E \subset \rD(H)\cap D$.
\item[\rm (2)] $\ov{\cT [\phi,\psi]}=\cT [\psi,\phi]$ for all $\phi,\psi\in D$. 
\item[\rm (3)] $HE\subset D_1$ and, for all $\psi, \phi\in E$, 
\begin{equation}
\cT [H\phi,\psi] -\ov{\cT [H\psi, \phi]}=-i(\phi,\psi).
\label{TH}
\end{equation}
\end{list}
Here $\ov{z}$ denotes the complex conjugate of $z$. 
We call $E$ the {ultra-weak CCR-domain} 
and $D$ the {symmetric domain} of $\cT $. 
\end{definition}

\subsection{New insights and main results}
In this paper we are concerned with time operators of Hamiltonians with purely discrete spectrum. 
We in particular focus on the one-dimensional harmonic oscillator. 
As far as we know, there has been huge amount of investigations of 
time operators of the one-dimensional harmonic oscillator,
there has been however few rigorous mathematical investigations. 
Not even the domains of time operators have been established. 
In addition, although the existence of several time operators of harmonic oscillators is known, we have never found any classification of them. Not only are there no rigorous results, but unfortunately there are also misleading results. No matter how much the algebraic relations of $p\f q$ and $qp\f$ are derived, we have the impression so that they are far from mathematics. It then appears to be an ad hoc study and 
we aspire to a more systematic research. Then the new insights of this paper are to investigate (1)-(6) below: 
(1) definition of $T_\eps$,
(2) representation of $T_\eps$, 
(3) extension of $T_\eps$, 
(4) analytic continuation of representation of $T_\eps$, 
(5) limit of $T_\eps$ as $\eps\to0$, 
(6) comparison of $T_\eps$ and $\TG$. 

The problems discussed and solved in this paper are listed below:
\begin{description}
\item[1. Definition of $\bm{\arctan t_\#}$.] We give the firm mathematical definitions of both 
$\arctan t_0$ and $\arctan t_1$, 
and discuss the domains of $\arctan t_0$ and $\arctan t_1$. 
We also show that both are unbounded operators in Theorems \ref{unbounded} and \ref{unbounded2}. 

\item[2. Representation of $\bm{\arctan t_\#}$.]
Let
$S_{\#,\eps}=-\frac{1}{\sqrt \eps}\arctan(\sqrt \eps t_\#)$ be given by 
\eqref{tay}. 
We introduce three subspaces: 
\begin{align}
 &\label{m1}\kM_{0,\eps}=\textrm{LH}\left\{x^{2n} e^{-\alpha x^2/(2\sqrt\eps)}\ \Big|\ \alpha\in(0,1), n\in\NN\right\},\\
 &\label{m0}\kM_{1,\eps}=\textrm{LH}\left\{x^{2n+1} e^{-\alpha x^2/(2\sqrt\eps)}\ \Big|\ \alpha\in(0,1), n\in\NN\right\},\\
\label{k}
 &\kK_\eps=\textrm{LH}\left\{x^{n} e^{-x^2/(2\sqrt\eps)}\ \Big|\ n\in \NN\right\}.
\end{align}
Here and in what follows $\textrm{LH}\{\ldots\}$ means the linear hull of $\{\ldots\}$ and $\NN=\{0,1,2,\ldots,\}$. 
\begin{description}
 \item[$(\alpha\in (0,1))$] 
 We describe the action 
 of $S_{\#,\eps}$ on $\kM_{\#,\eps}$ in 
Lemmas~\ref{main} and \ref{main2}:
 \begin{align*}
  S_{0,\eps} 
  x^{2n}e^{-\alpha x^2/(2\sqrt \eps)}
  &=-\frac{i}{2\sqrt \eps}\left(\lk x^2-2\sqrt\eps\frac{d}{d\alpha}\rk^n\AAA\right)e^{-\alpha x^2/(2\sqrt \eps)},\\
  S_{1,\eps}
  x^{2n+1}e^{-\alpha x^2/(2\sqrt \eps)}
  &=-\frac{i}{2\sqrt \eps}\left(\lk x^2-2\sqrt\eps\frac{d}{d\alpha}\rk^n\AAA\right)xe^{-\alpha x^2/(2\sqrt \eps)}.
 \end{align*}
 
 \item[$(\alpha=1)$]
 We show in Theorem \ref{domain} that 
 $$\rD(S_{\#,\eps})\cap\kK_\eps=\{0\},\quad \#=0,1$$ 
 and 
 it is also shown that any eigenvector of $h_\eps$ 
 does not belong to $\rD(S_{\#,\eps} )$.
 As a result we see that 
 $(f,S_{\#,\eps} g)$ diverges to infinity for any $f,g\in\kK_\eps$. 
\end{description}

\item[3. Ultra-weak time operator.]
We can see that 
$[h_\eps, S_{0,\eps}]=-i\one$ on $\kM_{0,\eps}$ and
$[h_\eps, S_{1,\eps} ]=-i\one$ on $\kM_{1,\eps}$, 
but $$\kM_{0,\eps}\cap \kM_{1,\eps}=\{0\}.$$ 
Let $\kM_\eps=\kM_{0,\eps}\oplus\kM_{1,\eps}$. 
We define the ultra-weak time operator 
$\cT _\eps $ associated with $T_\eps$. 
Note that $\cT_\eps$ is a sesqui-linear form:
$\cT_\eps\colon \kM_\eps\times\kM_\eps\to\CC$. 
We can see that 
$[h_\eps, \cT _\eps ]=-i\one$ holds true on 
$\kM_\eps$ in the sense of sesqui-linear forms in Theorem \ref{main3}.

\item[4. Matrix representations and analytic continuation]
Let $\eps=1$ and set $\cT_{\eps=1}=\cT$. 
We give the explicit form of 
$$a_{nm}=\cT[x^{n} e^{+iz x^2/2}, x^{m} e^{+iz x^2/2}]$$ for 
$z\in i(0,1)$ in Theorem \ref{fab}. 
This gives a matrix representation of~$\cT$. 
Since $z=i$ is a singular point of $\arctan z$ and 
$q\f p e^{-x^2/2}=ie^{-x^2/2}$, 
in Section~\ref{kuwata} we show that 
$\cT[x^{n} e^{+iz x^2/2}, x^{m} e^{+iz x^2/2}]$ diverges to infinity for $z=i$.
We discuss however the analytic continuation of the map 
$$z\mapsto \cT[x^{n} e^{+iz x^2/2}, x^{m} e^{+iz x^2/2}]$$ 
to $\HH\setminus\{i\alpha \mid \alpha\in[1,\infty)\}$ 
in Theorem \ref{anacon}, where
$\HH$ denotes the open upper half plane in~$\CC$. 

\item[5. Continuum limit.]
We are interested in the limit of $T_\eps$ as $\eps\to 0$ and deriving $T_{\rm AB}$ as the limit. 
By \eqref{T} we boldly hope that $T_\eps\to T_{\rm AB}$ as $\eps\to0$. Thus this is one reason to study the time operator of the form \eqref{T} 
in this paper. 
We shall consider this problem in the sense of sesqui-linear forms. 
We are interested in the limit of $\cT_\eps$ as $\eps\to 0$, 
and we expect that the limit is the sesqui-linear form $\cT_{\rm AB}$
associated with $T_{\rm AB}$, i.e., 
\begin{align}\label{cont limit}
\lim_{\eps\to 0}\cT _\eps =\cT_{\rm AB}
\end{align}
\eqref{cont limit} is called the continuum limit in this paper. 
We prove this in Theorem~\ref{cc}. 
\eqref{cont limit} is formally rewritten as 
\[
\lim_{\eps\to 0}
-\frac{1}{2\sqrt \eps}(\arctan (\sqrt \eps t_0) +\arctan (\sqrt \eps t_1)) 
=
-\half (t_0+t_1).
\]
This convergence gives a connection between 
time operators of self-adjoint operators with purely absolutely continuous spectrum and 
those of self-adjoint operators with 
purely discrete spectrum. 

\item[6. POVM.]
Let $\cT$ be the ultra-weak time operator $\cT _G$ associated with 
$\TG$. We show that $\cT\neq \cT _G$. 
This is shown in Theorem \ref{galapon2}. 
\end{description}

\subsection{Literatures}
\label{lit}
We introduce several literatures on angle operators and phase operators. Literatures listed below are mainly taken from physics. 
The canonical commutation relation and the uncertainty relation are found by W. Heisenberg \cite{hei27} at 1927. 
In 1945 the uncertain relation of time and energy is deduced from the principle of quantum mechanics in \cite{MT45}.

In 60's 
Aharonov and Bohm \cite{AB61} discussed the time operator for the case of a free quantum particle. The word \lq\lq time operator" appears in \cite{SG64} where 
a phase operator and an angle operator are studied. 
Also see \cite{raz67,CN68}. 
The mathematically rigorous study of $pq-qp=-i\one$ is given by \cite{fug67}. 
In \cite{ros69} time operators are studied on the set of distributions to avoid the difficulty of the domains of time operators. 
In \cite{lef69} a relationship between angle operators and phase operators is studied. 

In 80's it is shown in \cite{DD84} that no dense domains of canonical commutation relations exist under some conditions. 
Also in \cite{sch83a,sch83b} canonical commutation relations are studied from an operator theoretic point of view. 
In the series of papers \cite{BMP87,BD88,BD89a,BD89b} 
the algebraic properties of time operators of harmonic oscillators and related operators are studied, but they are far away from 
firm mathematics. 
In \cite{bau83} an extension of a Hilbert space is 
considered to define a self-adjoint phase operator. 
In \cite{GYS81a,GYS81b} 
$T_{\rm AB}$ is defined on the Schwartz space and 
$p^{-1}$ is given by $\lr{p^{-1}f}{ g}=\lr{ f}{ \int^x g(t) dt}$. 
 This is different from $p^{-1}$ defined in terms of the spectral measure of $p$. 
In \cite{PG89} the phase operator is discussed, but from only physical point of view. 

In 90's in \cite{CF91a,CF91b} time operators of 
Hamiltonians with purely discrete spectrum are studied. 
In particular \cite[(27b)]{CF91a} gives a time operator of 
a harmonic oscillator. 
In \cite{LLH96} 
a relationship between angle operators and phase operators is studied. 
We are inspired by papers \cite{LLH98,SV98} where the difficulty of the definition of the angle operator is pointed out. 

In the 21st century, 
in \cite{miy01}
time operators of Hamiltonians with purely 
absolutely continuous spectrum are given through the weak Weyl relations.
On the other hand, in the series of papers \cite{gal02a,gal02b,GCB04,GCB05,BG12}, 
time operators of Hamiltonians with purely 
discrete spectrum are studied. See also \cite{AM08,ara09}. 
In \cite{ter16} the existence of conjugate operators 
of Hamiltonians possessing purely point spectrum is established in an abstract setting. 
In \cite{AH17} the hierarchy of time operators are introduced, and 
a time operator of the hydrogen atom which has both continuous spectrum and discrete spectrum, is given.

\section{Definitions and properties of $\bm{S_{0,\eps}}$ and $\bm{S_{1,\eps}}$}
\subsection{Definitions of $\bm{S_{0,\eps}}$ and $\bm{S_{1,\eps}}$}
To consider time operators of $h_\eps$, 
it is useful to decompose $h_\eps$ with the even part and the odd part. 
Let 
\begin{align*}
&\LE=\{f\in\LR\mid \ f(-x)=f(x)\},\\
&\LO=\{f\in\LR\mid \ f(-x)=-f(x)\},
\end{align*}
i.e., $\LE$ denotes the set of all even $L^2$-functions, 
and $\LO$ all odd $L^2$-functions. 
We fix $0<\eps\leq 1$. 
We set 
$h_0=h_\eps\lceil_{\LE}$ and $h_1=h_\eps\lceil_{\LO}$. 
Since 
$h_\eps$ is reduced by $\LE$ and $\LO$, 
we have
\begin{align}\label{decomposition}
h_\eps=h_0\oplus h_1.
\end{align}
The spectrum of $h_\eps$ is
$\s(h_\eps)=\{\sqrt \eps(n+\half)\}_{n=0}^\infty$ and 
the eigenvector $e_n$ associated with the eigenvalue $\sqrt \eps(n+\half)$ with odd integer $n$ is an odd function, 
and that with even integer $n$ is an even function. 
They are given by 
 $e_n(x)=c_n H_n(x/\eps^{1/4}) e^{-x^2/(2\sqrt\eps)}$, where $H_n(x)$ is the $n$th Hermite polynomial given by 
\begin{align}\label{H}
H_n(x)=(-1)^ne^{x^2}\frac{d^n}{dx^n} e^{-x^2}=n!\sum_{k=0}^{[n/2]}\frac{(-1)^k}{k!}\frac{(2x)^{n-2k}}{(n-2k)!}
\end{align}
and $c_n$ is the normalized constant such that $\|e_n\|=1$. 
 $H_n$ is odd for odd integer $n$ and even for even integer $n$. 
 Set $\{c_n H_n(x/\eps^{1/4}) e^{-x^2/(2\sqrt\eps)}\mid n=2m\}$ is a complete orthonormal system of $\LE$, and 
 set $\{c_n H_n(x/\eps^{1/4}) e^{-x^2/(2\sqrt\eps)}\mid n=2m+1\}$ is that of $\LO$.
We have 
$\s(h_0)=\{\sqrt \eps(2m+\half)\}_{m=0}^\infty$ 
and 
$\s(h_1)=\{\sqrt\eps(2m+1+\half)\}_{m=0}^\infty$.

Since the operator $t=q^{-1}p$ maps $\rD(q^{-1}p)\cap L_0^2$ to $L_0^2$, we define the operator $S_{0,\eps}$ on $L_0^2$ by
\begin{align*}
 &\rD(S_{0,\eps} )=\left\{f\in L_0^2\cap\bigcap_{n=0}^\infty \rD(t_0^{2n+1})
 \ \bigg|\ 
 \text{s-}\!\!\!\!\lim_{N\to\infty} 
 \sum_{n=0}^N \frac{(-1)^n}{2n+1}(\sqrt\eps t_0 )^{2n+1}f\mbox{ exists}\right\},\\
 &S_{0,\eps} =-\frac{1}{\sqrt \eps}\sum_{n=0}^\infty \frac{(-1)^n}{2n+1}(\sqrt \eps t_0  )^{2n+1}.
\end{align*}
{\it Formally} we write $S_{0,\eps}$ as 
\begin{align}
\label{arctan}
S_{0,\eps} =-\frac{1}{\sqrt \eps}\arctan \sqrt\eps t_0 .
\end{align}
We give a remark on the expression \eqref{arctan}. 
The use of formal notation \eqref{arctan} is often useful since it enhances our imagination, but it can be seriously misleading. 
We need to be careful to use \eqref{arctan}. 

To study the properties of $t$ and $S_{0,\eps}$, we introduce two spaces below:
\begin{align*}
 &\LK_0=\textrm{LH}\left\{e^{-\alpha x^2/(2\sqrt \eps )}\ \middle|\ \alpha\in(0,1)\right\},\\
 &\LK_1=\textrm{LH}\left\{xe^{-\alpha x^2/(2\sqrt \eps )}\ \middle|\ \alpha\in(0,1)\right\}.
\end{align*}
\begin{lemma}
 The subspace $\LK_0+ \LK_1$ is dense in $\LR$, and 
 $\LK_0\perp \LK_1$. 
\end{lemma}
\proof
It is immediate to see that $\LK_0\perp \LK_1$. 
By the definition of $\LK_0$ we obtain that 
$$\frac{e^{-\alpha x^2/(2\sqrt \eps )}-e^{-\alpha' x^2/(2\sqrt \eps )}}{\alpha-\alpha'}\in 
\LK_0$$
for $\alpha,\alpha'\in(0,1)$.
In particular 
$$x^2e^{-\alpha x^2/(2\sqrt \eps )}=-2\sqrt \eps \lim_{\alpha'\to\alpha}
\frac{e^{-\alpha x^2/(2\sqrt \eps )}-e^{-\alpha' x^2/(2\sqrt \eps )}}{\alpha-\alpha'}\in 
\overline{\LK_0},$$
where $\overline{\LK_0}$ is the closure of $\LK_0$ in $L^2(\mathbb R)$.
Similarly we can see that 
$x^{2n}e^{-\alpha x^2/(2\sqrt \eps )}\in \overline{\LK_0}$ for $n\in\NN$.
In the same way we can also see that 
$x^{2n+1}e^{-\alpha x^2/(2\sqrt \eps )}\in \overline{\LK_1}$ for $n\in\NN$.
Since 
$\textrm{LH}\{x^n e^{-\alpha x^2/(2\sqrt \eps )}\mid n\in\NN\}$ is dense, 
the lemma follows. 
\qed

\begin{lemma}\label{L0}
The domain of the operator $S_{0,\eps}$ contains the subspace $\LK_0$. 
\end{lemma}
\proof
Let $\alpha\in(0,1)$ and $f(x)=e^{-\alpha x^2/(2\sqrt \eps )}\in \LK_0$. Then 
$\sqrt \eps t_0 f=i\alpha f$, i.e., $f$ is an eigenvector of $\sqrt \eps t_0 $ 
with eigenvalue $i\alpha $. 
Thus $(\sqrt \eps t_0 )^{2n+1} f =(-1)^n i \alpha^{2n+1} f$.
Hence 
\[\left\|\sum_{n=0}^\infty \frac{(-1)^n}{2n+1}(\sqrt\eps t_0 )^{2n+1}f\right\|\leq 
\|f\|\sum_{n=0}^\infty \frac{\alpha^{2n+1}}{2n+1} <\infty.\]
Then the lemma is proven.
\qed

Let $\s_{\rp}(A)$ be the set of all eigenvalues of $A$. 
 \begin{theorem}\label{unbounded}
The operators $t_0$ and $-S_{0,\eps}$ have eigenvalues $i\alpha$ for all $\alpha\in(0,\infty)$, i.e, 
$i (0,\infty)\subset \s_{\rp}(t_0)$ and $-i(0,\infty)\subset \s_{\rp}(S_{0,\eps} )$.
In particular $t_0$ and $S_{0,\eps} $ are unbounded. 
\end{theorem}
\proof
For $f(x)=e^{-\alpha x^2/(2\sqrt \eps )}$ with $\alpha\in(0,\infty)$ 
it follows that $\sqrt \eps t_0 f=i\alpha f$. Then the first statement follows. 
Let $\alpha\in(0,1)$. 
We have 
\begin{align*}
S_{0,\eps} f=
-\frac{1}{\sqrt\eps} 
\sum_{n=0}^\infty 
\frac{(-1)^n(i \alpha)^{2n+1}}{2n+1} f=
-\frac{i}{\sqrt\eps} \half \log\frac{1+ \alpha}{1- \alpha} f.
\end{align*}
Since $ \{\log\big((1+\alpha)/(1-\alpha)\big)\mid \alpha\in(0,1)\}=(0,\infty)$, 
the second statement is proven. 
\qed

Let us define $S_{1,\eps} $ in a similar way to $S_{0,\eps}$. 
Set 
\begin{align*}
 &\rD(S_{1,\eps})=\lkk f\in L_1^2\cap\bigcap_{n=0}^\infty \rD(t_1^{2n+1})
 \ \middle|\ 
 \text{s-}\!\!\!\!\lim_{N\to\infty}
 \sum_{n=0}^N \frac{(-1)^n}{2n+1}(\sqrt\eps t_1)^{2n+1}f\mbox{ exists}\rkk,\\
 &S_{1,\eps}=-\frac{1}{\sqrt\eps}
 \sum_{n=0}^\infty \frac{(-1)^n}{2n+1}(\sqrt \eps t_1)^{2n+1}.
\end{align*}
Formally we also write
\[
\label{arctanast}
S_{1,\eps} =-\frac{1}{\sqrt\eps} \arctan \sqrt\eps t_1.
\]

In a similar way to Lemma \ref{L0} and Theorem \ref{unbounded}, we can show the theorem below.
\begin{theorem}\label{unbounded2}
\begin{enumerate}[{\rm (1)}]
\item The domain of the operator $S_{1,\eps}$ contains the subspace $\LK_1$.
\item The operators $t_1$ and $-S_{1,\eps}$ have eigenvalues $i\alpha$ for all $\alpha\in(0,\infty)$, i.e, $i(0,\infty)\subset \s_{\rp}(t_1)$ and 
$-i(0,\infty)\subset \s_{\rp}(S_{1,\eps})$.
In particular $t_1$ and $S_{1,\eps} $ are unbounded. 
\end{enumerate}
\end{theorem}

\subsection{Representations of $\bm{S_{0,\eps}}$ and $\bm{S_{1,\eps}}$ for $\bm{\alpha\in (0,1)}$}
\label{representation}
To define a time operator of $h_\eps$, we prepare several technical lemmas 
on the properties of $S_{\#,\eps}$. 
\begin{lemma}\label{key}
 Suppose that $\alpha\in (0,1)$. 
 Then, for all $m\in\mathbb N$, $x^{2m}e^{-\alpha x^2/(2\sqrt \eps)}\in\mathrm D(S_{0,\eps})$ and 
 \begin{align*}
  S_{0,\eps} 
  x^{2m}e^{-\alpha x^2/(2\sqrt \eps)}
  =
  -\frac{i}{2\sqrt \eps}\sum_{k=0}^m \binom{m}{k} (-2\sqrt\eps)^k 
  \left(
  \log\frac{1+\alpha}{1-\alpha}\right)^{(k)}x^{2m-2k}e^{-\alpha x^2/(2\sqrt \eps)}.
 \end{align*}
 Here $F^{(k)}=d^k F/d\alpha^k$. 
\end{lemma}
\proof
Let $f(x)=e^{-\alpha x^2/(2\sqrt \eps)}$.
We can compute $(\sqrt\eps t_0 )^n x^{2m}f(x)$ directly as 
\[
(\sqrt\eps t_0 )^n x^{2m}f(x)=
\sum_{k=0}^m \binom{n}{k}
\binom{m}{k}k!2^kx^{2m-2k}(i\alpha )^{n-k}(-i\sqrt\eps)^kf(x).\] 
From this, we see that 
\begin{align}
 &\sum_{n=0}^M\frac{(-1)^n}{2n+1}(\sqrt\eps t_0 )^{2n+1}x^{2m}f(x)\nonumber \\
 &=
 \sum_{n=0}^M\frac{(-1)^n}{2n+1}
 \sum_{k=0}^m \binom{2n+1}{k}\binom{m}{k}k!2^kx^{2m-2k}(i\alpha )^{2n+1-k}(-i\sqrt\eps)^kf(x)\nonumber \\
 &\label{aki}
 =
 i
 \sum_{n=0}^M\left(
 \frac{\alpha^{2n+1}}{2n+1}x^{2m}
 -2m\sqrt\eps \alpha^{2n}x^{2m-2}
 +\sum_{k=2}^m \binom{m}{k} (-2\sqrt\eps)^k \left(\alpha^{2n}\right)^{(k-1)}
 x^{2m-2k}\right)f(x).
\end{align}
Hence we can see that as $M\to\infty$, 
\begin{align*}
 &\sum_{n=0}^M\frac{(-1)^n}{2n+1}t_0^{2n+1}x^{2m}f(x)\\
 &\to
 i\left(
 \half\log\frac{1+\alpha}{1-\alpha} 
 x^{2m}- \frac{2m\sqrt\eps}{1-\alpha^2} x^{2m-2}
 +\sum_{k=2}^m \binom{m}{k} (-2\sqrt\eps)^k 
 \left(
 \frac{1}{1-\alpha^2}\right)^{(k-1)}x^{2m-2k}\right)f(x)\\
 &=
 \frac{i}{2}\sum_{k=0}^m \binom{m}{k} (-2\sqrt\eps)^k 
 \left(
 \log\frac{1+\alpha}{1-\alpha}\right)^{(k)}x^{2m-2k}f(x).
\end{align*}
Then for each $x\in\RR$, 
\begin{align}
\label{aki2}
\lim_{M\to\infty}
\sum_{n=0}^M\frac{(-1)^n}{2n+1}t_0^{2n+1}x^{2m}f(x)
=\frac{i}{2}\sum_{k=0}^m \binom{m}{k} (-2\sqrt\eps)^k 
 \left(
 \log\frac{1+\alpha}{1-\alpha}\right)^{(k)}x^{2m-2k}f(x).
 \end{align}
By \kak{aki}, 
$\sum_{n=0}^M\frac{(-1)^n}{2n+1}t_0^{2n+1}x^{2m}f(x)$ is 
a $2m$-degree polynomial multiplied by $f$.
Moreover the coefficient $c_{2m-2k, \alpha, M}$ 
of $x^{2m-2k}$ of the polynomial is bounded as 
$|c_{2m-2k, \alpha, M}|\leq c_{2m-2k, \alpha}$ with some constat 
$c_{2m-2k, \alpha}$. Then the Lebesgue dominated convergence theorem yields that \kak{aki2} is in the sense of $L^2$. 
 Then the lemma is proven. 
 \qed

Let us set 
\begin{equation*}
 \xae=e^{-\alpha x^2/(2\sqrt\eps)},\quad \taxe =2\lk \frac{x^2}{2}-\sqrt\eps\frac{d}{d\alpha}\rk. \label{tr}
\end{equation*}
\begin{lemma}\label{main}
 Let $\alpha\in (0,1)$ and $f_{2m}=x^{2m}\xae$ for $m\in\NN$. 
 Then 
 \begin{align}
  S_{0,\eps} f_{2m}=
  -\frac{i}{2\sqrt \eps}\lk \taxe^m \AAA \rk 
  \xae .
 \end{align}
 In particular, 
 the matrix representation of $S_{0,\eps}$ is given by 
 \begin{align}
  (f_{2n}, S_{0,\eps} f_{2m})=-\frac{i}{2\sqrt \eps}
  \lk\xae, x^{2n}\!
  \lk \taxe ^m \AAA \rk 
  \xae \rk. 
 \end{align}
\end{lemma}
\proof
By Lemma \ref{key}, we have
\begin{align*}
 S_{0,\eps} f_{2m} 
 &=
 -\frac{i}{2\sqrt \eps}\sum_{k=0}^m \binom{m}{k} (-2\sqrt\eps)^k 
 \left(\log\frac{1+\alpha}{1-\alpha}\right)^{(k)} x^{2m-2k}\xa\\ 
 &=
 -\frac{i}{2\sqrt \eps}
 \lk \taxe ^m \AAA \rk \xae.
\end{align*}
Thus the lemma is proven. 
\qed
In the same way as in the proof of Lemma \ref{key}, we can show that, for all $m\in\mathbb N$, 
\begin{equation*}
 x^{2m+1}e^{-\alpha x^2/(2\sqrt \eps)}\in\mathrm D(S_{1,\eps})
\end{equation*}
and we have the representation of 
$S_1$ 
as in Lemma \ref{main}. 

\begin{lemma}\label{main2}
 Suppose that $\alpha\in (0,1)$. 
 Set $f_{2m+1}=x^{2m+1}\xae$ for $m\in\NN$. 
 Then \begin{align}
  S_{1,\eps} f_{2m+1}=
  -\frac{i}{2\sqrt \eps}\lk \taxe ^m \AAA \rk 
  x\xae .
 \end{align}
 In particular, 
 the matrix representation of $S_{0,\eps}$ is given by 
 \begin{align}
  (f_{2n+1}, S_{1,\eps} f_{2m+1})=
  -\frac{i}{2\sqrt \eps}\lk
  \xae, x^{2n+2}\lk \taxe ^m \AAA \rk \xae\rk. 
 \end{align}
\end{lemma}
\proof
The proof is similar to that of Lemma \ref{main}.
\qed
We have a brief comment on the derivation of representations of $S_{0,\eps}$ and $S_{1,\eps}$. 
These representations can be formally derived from a generating-function-type argument. 
Note that 
\begin{equation*}
 \arctan(i\alpha )=\frac{i}{2}\log\frac{1+\alpha}{1-\alpha}=
 i\int \frac{1}{1-\alpha^2}d\alpha.
\end{equation*} 
Since 
$-\sqrt \eps S_{0,\eps}\xa =\arctan(i\alpha )\xae$, taking the derivative on $\alpha$ on both sides we obtain that 
\[-\sqrt \eps S_{0,\eps}\lk -\frac{x^2}{2\sqrt\eps}\rk \xae=\lk \frac{d}{d\alpha}\arctan(i\alpha )\rk \xae+
\arctan(i\alpha )\lk -\frac{x^2}{2\sqrt\eps}\rk\xae.\]
Then we have 
\[S_{0,\eps} x^2\xae=-\frac{2}{\sqrt \eps}\lkk 
\lk\frac{x^2}{2}-\sqrt\eps\frac{d}{d\alpha}\rk \arctan(i\alpha )
\rkk \xae.\]
Repeating this we can see Lemma \ref{main}. 
The following statement is immediate. 
\begin{lemma}\label{r}
 Suppose that $\alpha\in (0,1)$. 
 Let $\rho$ be a polynomial. 
 Then 
 \begin{align}
  \label{r1}S_{0,\eps}\rho(x^2)\xae &=-\frac{i}{2\sqrt \eps} \lk \rho(\taxe) \AAA\rk \xae,\\
  \label{r2}S_{1,\eps} \rho(x^2)x\xae &=-\frac{i}{2\sqrt \eps} \lk \rho(\taxe) \AAA\rk x\xae.
 \end{align}
\end{lemma}

Although Lemma \ref{r} may be extended to general functions $\rho$, 
we do not discuss it here, but 
give an example. 
\begin{example}
 Let $\rho(x^2)=e^{+\beta x^2/(2\sqrt\eps)}$ with $0<\alpha-\beta<1$. 
 Then 
 \[ \rho(\tax)=e^{\beta \tax /(2\sqrt\eps)}=e^{\beta x^2/(2\sqrt\eps)}e^{-\beta\frac{d}{d\alpha}}.\] 
 Thus 
 \[
 e^{\beta \tax /(2\sqrt\eps)}
 \log\frac{1+\alpha}{1-\alpha}
 =
 e^{\beta x^2/(2\sqrt\eps)}
 \log\frac{1+(\alpha-\beta) }{1-(\alpha-\beta)},\]
 and hence 
 \begin{align*}
 S_{0,\eps} e^{-(\alpha-\beta)x^2/(2\sqrt\eps)}&=
 -\frac{i}{2\sqrt \eps} 
 \lk
 e^{\beta \tax /(2\sqrt\eps)}
 \log\frac{1+\alpha }{1-\alpha}\rk
 e^{-\alpha x^2/(2\sqrt\eps)}\\
 &=
 -\frac{i}{2\sqrt \eps} 
 \log\frac{1+(\alpha-\beta) }{1-(\alpha-\beta)}e^{-(\alpha-\beta) x^2/(2\sqrt\eps)}.
 \end{align*}
\end{example}

\subsection{Representation of $\bm{S_{0,\eps}}$ and $\bm{S_{1,\eps}}$ for 
$\bm{\alpha=1}$}
\label{kuwata}
In this section, we discuss the case of $\alpha=1$. 
We see that, for $\alpha\in(0,1)$,
\[S_{0,\eps} e^{-\alpha x^2/(2\sqrt\eps)}=
-\frac{i}{2\sqrt \eps} \log \frac{1+\alpha}{1-\alpha}e^{-\alpha x^2/(2\sqrt\eps)}.\]
We also see that 
\[S_{0,\eps} x^{2}e^{-\alpha x^2/(2\sqrt\eps)}=
-\left(\frac{i}{2\sqrt \eps} \log \frac{1+\alpha}{1-\alpha}x^2
-i\frac{2}{1-\alpha^2}\right)e^{-\alpha x^2/(2\sqrt\eps)}.\]
Hence the coefficient of the leading term in the curly bracket diverges to infinity as $\alpha\to 1$. Then it is expected that 
$e^{-x^2/(2\sqrt\eps)}, x^{2}e^{-x^2/(2\sqrt\eps)}\not\in \rD(S_{0,\eps})$. Let $\kK_\eps$ be \eqref{k}. 
\begin{theorem}\label{domain}
For $\#=0,1$, $\kK_\eps\cap \rD(S_{\#,\eps})=\{0\}$. 
\end{theorem}
\proof
We consider only $S_{0,\eps}$. 
Let $\rho_1(x)$ be a polynomial with $\rho_1(x)\neq\rho_1(-x)$. 
We see that $x^{2m+1}e^{-x^2/(2\sqrt\eps)}\not\in \rD((\sqrt\eps t_0 )^{m+1})$ and hence 
$\rho_1(x)e^{-x^2/(2\sqrt\eps)}\not\in \rD(S_{0,\eps})$. 

Let $\rho_0(x)$ be a monic polynomial of degree $2m$ such that $\rho_0(x)=\rho_0(-x)$. Set
$\xbe=e^{-x^2/(2\sqrt\eps)}$. We have 
\begin{align}
 &\sum_{n=0}^M\frac{(-1)^n}{2n+1}(\sqrt\eps t_0 )^{2n+1}\rho_0(x)\xbe\nonumber\\
 &=
 i
  \sum_{n=0}^M\left(
 \frac{x^{2m}}{2n+1}
 -
 \frac{\sqrt\eps\rho_0'(x)}{x} 
 +
 \sum_{k=2}^m \frac{(-1)^k}{k!} 
 2n(2n-1)\cdots(2n-k+2)
 \left((\sqrt\eps t_0 )^k\rho_0(x)\right)\right)\xbe.\label{hiro}
\end{align}
Let $e_{2n}$ be the eigenvector of $h_\eps$ with eigenvalue $2n$. We have 
\[\biggl\{\left((\sqrt\eps t_0 )^k\rho_0(x)\right)\xbe\ \bigg|\ 1\leq k\leq m\biggr\}\subset \textrm{LH}\{e_0,e_2,\ldots, e_{2m-2}\}\]
and henceforth $\big\{\left((\sqrt\eps t_0 )^k\rho_0(x)\right)\xbe\mid 1\leq k\leq m\big\}\perp \{e_{2m}\}$. 
From \eqref{hiro} it follows that 
\begin{align*}
 \left|\lk 
 e_{2m}, 
 \sum_{n=0}^M\frac{(-1)^n}{2n+1}(\sqrt\eps t_0 )^{2n+1}\rho_0(x)\xbe\rk\right|
 &=
 \left|
 \sum_{n=0}^M\lk 
 e_{2m}, 
 \frac{1}{2n+1}x^{2m}
 \xbe\rk\right|\\
 &=
 |\lk 
 e_{2m}, x^{2m}
 \xbe\rk|\sum_{n=0}^M
 \frac{1}{2n+1}\to\infty\ \ (M\to\infty).
\end{align*} 
Hence 
$\sum_{n=0}^M\frac{(-1)^n}{2n+1}(\sqrt\eps t_0 )^{2n+1}\rho_0(x)\xbe$ does not 
converge as $M\to\infty$ and 
$\rho_0(x)\xbe\not\in \rD(S_{0,\eps})$. 
\qed

\begin{corollary}\label{domain2}
Let $e_n$ be the $n$th eigenvector of $h_\eps$. Then $e_n\not\in \rD(S_{\#,\eps})$ for $\#=0,1$. 
\end{corollary}
\proof
Since $e_n\in\mathfrak{K}_\eps$, from Theorem \ref{domain}, it follows that $e_n\not\in \rD(S_{\#,\eps})$.
\qed

\section{Matrix representations}
\subsection{Ultra-weak time operators}
We firstly confirm that $S_{\#,\eps}$ is a conjugate operator of $h_\eps$. 
Let $\kM_{0,\eps}$ and $\kM_{1,\eps}$ be \eqref{m1} and \eqref{m0}, respectively. 
\begin{lemma}\label{s}
For $\#=0,1$, 
$[h_\eps,S_{\#,\eps} ]=-i\one$ on $\kM_{\#,\eps}$. 
\end{lemma}
\proof
We shall show 
$[h_\eps,S_{0,\eps}]=-i\one$ on $\kM_{0,\eps}$. 
It is sufficient to show that, for $n\in\mathbb N$, $\alpha\in(0,1)$ and $\xae =e^{-\alpha x^2/(2\sqrt\eps)}$,
\begin{equation*}
[h_\eps, S_{0,\eps}]x^{2n}\xae=-ix^{2n}\xae.
\end{equation*}
From Lemma \ref{key}, we see that
\begin{align*}
 &\sqrt \eps h_\eps S_{0,\eps} x^{2n}\xae\\
 &=\frac{i}{4}\left(\lkk
 \bigl((1-\alpha^2)x^2+\alpha\bigr)\taxe^{n}-2n(1-2\alpha x^2)\taxe^{n-1}-4n(n-1)x^2\taxe^{n-2}\rkk\log\left(\frac{1+\alpha}{1-\alpha}\right)\right)\xae,\\
 &\sqrt \eps S_{0,\eps} h_\eps x^{2n}\xae\\
 &=\frac{i}{4}\left(\lkk
 (1-\alpha^2)\taxe^{n+1}+(4n+1)\alpha \taxe^n-2n(2n-1)\taxe^{n-1}\rkk\log\left(\frac{1+\alpha}{1-\alpha}\right)\right)\xae.
\end{align*}
Therefore, by Lemma \ref{a1}, \eqref{K1} and \eqref{K2} below we have
$[h_\eps, S_{0,\eps}]x^{2n}\xae=-ix^{2n}\xae$. 
In a similar way to this 
 it follows that 
 $[h_\eps, S_{1,\eps}]=-i\one$ on $\kM_{1,\eps}$.
\qed

Intuitively we expect that $$T_\eps=\half (S_{0,\eps} +S_{1,\eps})$$ is a time operator 
of $h_\eps$. 
We notice that 
$\kM_{0,\eps}\subset \LE$ and 
$\kM_{1,\eps}\subset \LO$. 
Furthermore
\begin{align*}
t_0^n \colon \kM_{0,\eps}\to \LE,\quad 
t_1^n \colon \kM_{1,\eps}\to \LO,
\end{align*}
but 
$\rD(t_0^n)\cap \kM_{1,\eps}=\{0\}$
and 
$\rD(t_1^n)\cap \kM_{0,\eps}=\{0\}$.
From this it is not obvious 
to find a dense domain of $T_\eps$ 
and 
to construct a dense domain $D$ such that 
$[h_\eps, T_\eps]=-i\one$ on $D$. 
So, instead of considering time operators, we define an ultra-weak time operator of $h_\eps$. 
By the decomposition 
$h_\eps=h_0\oplus h_1$ 
it is enough to construct ultra-weak time operators 
$\cT _0$ and $\cT _1$ of $h_0$ and $h_1$, respectively. 
Hence $\cT _0\oplus \cT _1$ becomes an ultra-weak time operator of $h_\eps$. 
We define 
\begin{align}
\cT_{0,\eps}[\psi,\phi]&=\half 
((\psi, S_{0,\eps} \phi)+(S_{0,\eps} \psi,\phi)),\quad \psi,\phi\in\rD(S_{0,\eps}),\\
\cT_{1,\eps}[\psi,\phi]&=\half 
((\psi, S_{1,\eps} \phi)+(S_{1,\eps} \psi,\phi)),\quad 
\psi,\phi\in\rD(S_{1,\eps}).
\end{align}

\begin{lemma}\label{ss}
The sesqui-linear form $\cT_{0,\eps}$ (resp. $\cT_{1,\eps}$) 
is an ultra-weak time operator of $h_0$ (resp. $h_1$) 
with the symmetric domain $\rD(S_{0,\eps})$ (resp. $\rD(S_{1,\eps})$) and 
the ultra-weak CCR-domain $\kM_{0,\eps}$ (resp. $\kM_{1,\eps}$). 
\end{lemma}
\proof
From Lemma \ref{s} it follows that 
$\cT_{0,\eps}[h_0\psi,\phi]-
\cT_{0,\eps}[\psi,h_0\phi]=-i(\psi,\phi)$ for $\psi,\phi\in\kM_{0,\eps}$ 
 and 
$\cT_{1,\eps}[h_1\psi,\phi]-
\cT_{1,\eps}[\psi,h_1\phi]=-i(\psi,\phi)$ for $\psi,\phi\in\kM_{1,\eps}$. 
The the lemma is proven. 
\qed
Let 
$\kM_\eps=\kM_{0,\eps}\oplus \kM_{1,\eps}$ and 
$\kD_\eps=\rD(S_{0,\eps})\oplus\rD(S_{1,\eps})$. 
The sesqui-linear form $\cT _\eps$ is defined by 
\[\cT_\eps=\cT_{0,\eps}\oplus \cT_{1,\eps}\]
with the domain $\kD_\eps\times \kD_\eps$.
We see that 
\[
\cT _\eps[\psi_0\oplus\psi_1,\phi_0\oplus \phi_1]
=
\cT_{0,\eps}[\psi_0,\phi_0]
+
\cT_{1,\eps}[\psi_1,\phi_1]\]
for $\psi_0\oplus\psi_1,\phi_0\oplus \phi_1\in \kD_\eps$.

\begin{theorem}\label{main3}
The sesqui-linear form $\cT _\eps$ is an ultra-weak time operator of
$h_\eps$ under the decomposition 
$h_\eps=h_0\oplus h_1$ with the symmetric domain $\kD_\eps$ and 
the ultra-weak CCR-domain~$\kM_\eps$. 
\end{theorem}
\proof
Let $\phi=\phi_0\oplus \phi_1, \psi=\psi_0\oplus \psi_1\in \kM_\eps$. 
By Lemma \ref{ss}, we have 
\begin{align*}
\cT _\eps[h_\eps\phi,\psi]-
\cT _\eps[\phi,h_\eps\psi]
&=
\cT_{0,\eps}[h_0\phi_0,\psi_0]-
\cT_{0,\eps}[\phi_0,h_0\psi_0]
+
\cT_{1,\eps}[h_1\phi_1,\psi_1]-
\cT_{1,\eps}[\phi_1,h_1\psi_1]\\
&=
-i(\phi_0,\psi_0)-i(\phi_1,\psi_1)\\
&=-i(\phi,\psi).
\end{align*}
Then the theorem is proven. 
\qed

\subsection{Matrix representations of $\bm{\cT_\eps}$ for $\bm{0<\alpha<1}$}
We are interested in computing $\cT_\eps [f,g]$ for $f,g\in \kM_\eps$. 
More precisely we want to see the explicit form of the function $K_{ab}$ such that 
\begin{align}\label{knm}
 \cT _\eps[x^a\xae, x^b \xce]=(\xae, K_{ab}\xce),\quad a,b\in\NN,\ \alpha,\beta\in(0, 1).
\end{align}

\begin{theorem}\label{fab}
 Fix $\alpha,\beta\in (0,1)$. 
 Let $f_a=x^a\xae$ and $f_b=x^b\xce$. 
 Then 
 \begin{align*}
  \cT_\eps [f_a, f_b]=
  \begin{cases}
   \displaystyle 
   \frac{i}{4\sqrt \eps} 
   \lk \xae , \lkk x^{2m}\taxe^n \AAA-x^{2n}\tbxe^m\BBB\rkk \xce \rk
   & \text{$a=2n$\phantom{,,}}\atop\text{$b=2m$,}\vspace{5pt}\\
   \displaystyle 
   \frac{i}{4\sqrt \eps} 
   \lk \xae , \lkk x^{2m+2}\taxe^n\AAA -x^{2n+2}\tbxe^m\BBB \rkk \xce \rk
   & \text{$a=2n+1$\phantom{,,}}\atop\text{$b=2m+1$,}\vspace{5pt}\\
   \displaystyle 
   \ 0 & \text{otherwise}.
  \end{cases}
 \end{align*}
 \end{theorem}
\proof
If $a+b$ is odd, 
then 
$\cT _\eps[f_a, f_b]=0$ by the definition of 
$\cT _\eps$. 
If $a+b$ is even, then 
$\cT _\eps[f_a, f_b]=\half \lkk (S_0f_a,f_b)+(f_a,S_0f_b)\rkk$ for $a=2n, b=2m$ and 
$\cT _\eps[f_a, f_b]=\half\lkk (S_1 f_a,f_b)+(f_a,S_1 f_b)\rkk$ for $a=2n+1, b=2m+1$. Then 
the theorem follows from Lemmas \ref{main} and \ref{main2}.
\qed

We can also see the corollary below. 
\begin{corollary}\label{lab}
 Fix $\alpha,\beta\in (0,1)$. 
 Let $H_a$ be the $a$th Hermite polynomial given by \eqref{H}. 
 Let $f_a=H_a\xae$ and $f_b=H_b\xce$. 
 Then $\cT_\eps [f_a, f_b]=(\xae, L_{ab} \xce)$, where 
 \begin{align*}
  L_{ab}
  =
  \begin{cases}
   \displaystyle 
   \frac{i}{4\sqrt \eps} 
   \left(H_{2m}(x)h_{n}(\taxe^n) \AAA-H_{2n}(x)h_{m}(\tbxe^m)\BBB\right)
   & \text{$a=2n$\phantom{,,}}\atop\text{$b=2m$,}\vspace{5pt}\\
   \displaystyle 
   \frac{i}{4\sqrt \eps}x 
   \left(H_{2m+1}(x)k_{n}(\tax ^n)\AAA-H_{2n+1}(x)k_{m}(\tbxe^m)\BBB\right)
   & \text{$a=2n+1$\phantom{,,}}\atop\text{$b=2m+1$,}\vspace{5pt}\\
   \displaystyle 
   \ 0& \text{otherwise}.
  \end{cases}
 \end{align*}
 Here $h_{n}$ and $k_{n}$ are defined by 
 $H_{2n}(x)=h_{n}(x^2)$ and $H_{2n+1}(x)=k_{n}(x^2) x$.
\end{corollary}

\begin{remark}
We give a comment on a matrix representation for $\alpha=1$. 
The matrix representation of $\cT_\eps $ with respect to $\{e_n\}$ is formally defined by $\cT_\eps [e_n, e_m]$. 
However $\cT_\eps [e_n, e_m]$ diverges to infinity 
by Corollary \ref{domain2}. 
Nevertheless in \cite{BG12} a matrix representation on $\{e_n\}$ is discussed through an analytic continuation. 
\end{remark}

\subsection{Analytic continuations of matrix representations}
\label{cayley}
In this section, we set $\eps=1$. 
We denote $\xi_{i\alpha ,\eps=1}=\xa$, $\cT _{\eps=1}=\cT $, 
$S_{\#,\eps=1}=S_\#$ and $h_{\eps=1}=h$ for notational simplicity. 
In Section \ref{representation} we show that 
$S\rho(x^2)\xa=\rho(\tax)\arctan(i\alpha ) \xa$ for $\alpha\in (0,1)$. 
We can extend $i\alpha $
 to $z\in \CC$. 
Let $$\xz=e^{+izx^2/2}\quad z\in\CC.$$
$\xz\in \LR$ if and only if $\Im z>0$. 
We also have $t_0\xz=z\xz$ for $z\in\CC$. 
Let $\cC \colon \CC\setminus\{-i\}\to\CC$ be the Cayley transform defined by 
$$\cC (z)=\frac{z-i}{z+i}.$$
We sometimes omit the variable $z$ and write $\cC (z)$ simply as $\cC $. 
In terms of the Cayley transform we have 
 $$\log\frac{1+\alpha}{1-\alpha}=-\log(-\cC (i\alpha )).$$ 
In particular, 
 $2\arctan z=-i\log (-\cC (z))$ follows. 
 Here and in what follows we define 
 $$\log z=\log|z| +i \arg z,\quad 
 -\pi \leq \arg z<\pi.$$
Hence $\log z$ is analytic on $\CC\setminus(-\infty, 0]$. 
 Let $$\HH=\{z\in \CC\mid \Im\,z>0\}$$ be the open upper half plane 
 and $$\DD =\{z\in\CC\mid |z|<1\}$$ be the open unit disc in $\CC$. 
The Cayley transform maps as 
$\cC(\HH)=\DD$, $\cC(\RR)=\partial \DD$,
$\cC(i[1,\infty))=[0,1)$,
$\cC(i[0,1))=[-1,0)$, 
$\cC(i(-1,0])=(-\infty,-1]$ and 
$\cC(i(-\infty,-1))=(1,\infty)$, 
where $\partial \DD $ denotes the unit circle in $\CC$. 
 Thus it is known that $z\mapsto -\log(-\cC (z))$ is analytic on 
 $\CC\setminus \{i\alpha \mid \alpha\in(-\infty,-1]\cup [1,\infty)\}$. 
Let 
\begin{align*}
&\kN_0=\textrm{LH}\{\rho(x^2)\xz\mid \rho\text{ is a polynomial, }z\in \HH\setminus\{i\}\},\\
&\kN_1=\textrm{LH}\{\rho(x^2)x\xz\mid \rho\text{ is a polynomial, }z\in \HH\setminus\{i\}\},\\
&\kN=\kN_0\oplus\kN_1.
\end{align*}
We also define $\tzx$ by 
$$\tzx=2\lk \frac{x^2}{2}-i\frac{d}{dz}\rk .$$
Now we define $\tilde S_\#$ for $\#=0,1$ by 
\begin{align*}
&\tilde S_0 \rho(x^2)\xz=\frac{i}{2}\lk \rho(\tzx)\log(-\cC )\rk \xz,\\
&\tilde S_1 \rho(x^2)x\xz=\frac{i}{2}\lk \rho(\tzx)\log(-\cC )\rk x\xz.
\end{align*}
We see that, 
for $z=i\alpha $ and $\alpha\in (0,1)$, 
the actions of $\tilde S_\#$ and $S_\#$ on $\rho(x^2)\xa$ are identical:
\[\tilde S_\#\rho(x^2)\xa=S_\# \rho(x^2)\xa.\] 

\begin{lemma}\label{a1}
For $\#=0,1$, $[h, \tilde S_\#]=-i\one $ on $\kN_\#$. 
\end{lemma}
\proof
Here we summarise formulae used in computations. 
\begin{enumerate}[(1)]
\item
$\displaystyle \tzx \log(-\cC )=x^2\log(-\cC )+\frac{4}{1+z^2}$,
\item
$[h,\tzx]=-1-2ixp$,\ \ 
$\bigl[[h,\tzx],\tzx\bigr]=-4x^2$,
\item
$\displaystyle [h,\tzx^n]=n\tzx^{n-1}(-1-2ixp)+ \half n(n-1) \tzx^{n-2}(-4x^2)$,
\item
$\displaystyle hx^{2n}\xz=
\half\bigl((1+z^2)x^{2n+2}-i(4n+1)zx^{2n}-2n(2n-1)x^{2n-2}\bigr)\xz$.
\end{enumerate}
We define the complex differential operator 
$Q_n$ by 
\begin{align}
Q_n=\lkk
\begin{array}{ll}
\displaystyle 1+z^2& n=0, \\
\displaystyle (1+z^2)\tzx-i4 z& n=1, \\
\displaystyle (1+z^2)\tzx^n-i4nz\tzx^{n-1}-4n(n-1)\tzx^{n-2}& n\geq2. 
\end{array}\right.\end{align}
Note that above formulae are valid on regular analytic functions on $\CC$. 
Using formulae (1)-(4) and 
$\tilde S_0x^{2n}\xi=-\frac{i}{2} \tzx^n \log(-\cC)\xz$, 
we can directly see that 
\begin{align}
 &\label{K1}h\tilde S_0x^{2n}\xz=\frac{i}{4}\Bigl(\lkk
 (1+z^2)x^2\tzx^{n}-iz \tzx^n-2n(1+2izx^2)\tzx^{n-1}-4n(n-1)x^2\tzx^{n-2}\rkk\log(-\cC )\Bigr)\xz,\\
 &\label{K2}\tilde S_0hx^{2n}\xz=\frac{i}{4}\Bigl(\lkk
 (1+z^2)\tzx^{n+1}-i(4n+1)z\tzx^n-2n(2n-1)\tzx^{n-1}\rkk\log(-\cC )\Bigr)\xz.
\end{align}
Note that 
$\tzx^{m}\log(-\cC )=
x^2 \tzx^{m-1} \log(-\cC )+\tzx^{m-1}\frac{4}{1+z^2}$, $m=n+1,n,n-1$.
Inserting 
these identities into the right hand side of 
$\tilde S_0hx^{2n}\xz$, 
we can see that 
\begin{align}
[h, \tilde S_0]x^{2n}\xz=-i\lk Q_n\frac{1}{1+z^2}\rk \xz\quad n\geq0.
\end{align}
The similar identity is obtained for $\tilde S_1$:
\begin{align}
[h, \tilde S_1]x^{2n+1}\xz=-i\lk Q_n\frac{1}{1+z^2}\rk x\xz\quad n\geq0.
\end{align}
Next we shall show that 
\begin{align}\label{hitoto}
(Q_nF)(z)=\tzx^n (1+z^2)F(z)
\end{align}
 for any regular analytic function $F$ on $\CC$. 
For $n=0,1$ one can show \eqref{hitoto} directly. 
Note that $[\tzx, z]=-2i$ and $[\tzx, z^2]=-4iz$.
Thus we have 
\[(1+z^2) \tzx^nF(z)=
\left(\tzx^n(1+z^2)+ 4izn \tzx^{n-1}+4n(n-1)\tzx^{n-2}\right)F(z).\]
Then \eqref{hitoto} follows. 
Together with them 
we have 
\[[h,\tilde S_0]x^{2n}\xz=-i\lk \tzx^n (1+z^2)\frac{1}{1+z^2}\rk \xz=
-i\lk \tzx^n 1\rk \xz
=
-ix^{2n} \xz\]
 and similarly 
$[h,\tilde S_1]x^{2n+1}\xz=-ix^{2n+1}\xz$. 
Then the lemma follows. 
\qed

Let $G_0(i\alpha )=(f, S \rho(x^2)\xa)$ for $f\in\LR$. 
The function $G_0$ can be extended to $z\in \DD \cap \HH$. 
Actually we can see that $\rho(x^2)\xz\in \rD(\tilde S_0)$ and 
\begin{align}
\label{g}
G_0(z)=\left(f, \tilde S_0 \rho(x^2)\xz\right)=\left(f, \frac{i}{2}
\lk \rho(\tzx)\log(-\cC)\rk \xz\right),\quad z\in \DD \cap \HH.
\end{align}
This can be proven in a similar way to Lemma \ref{r}. 
\begin{lemma}\label{b2}
Let $f\in\LR$. Define 
$F_0(z)=\bigl(f, \tilde S_0 \rho(x^2)\xz\bigr)$ for $z\in \HH\setminus\{i\}$. 
Then the function 
$F_0$ is analytic on $\HH\setminus\{i\alpha \mid \alpha\in[1,\infty)\}$. 
In particular 
$F_0$ is the analytic continuation of $G_0$ given by \eqref{g} 
to $\HH\setminus\{i\alpha \mid \alpha\in[1,\infty)\}$. 
\end{lemma}
\proof
We see that 
$F_0$ is differentiable on $z\in \HH\setminus\{i\alpha \mid \alpha\in[1,\infty)\}$ with 
\begin{align*}
\frac{d}{dz}F_0(z)=
\lk 
f, \frac{i}{2}\rho(\tzx)\frac{d}{dz}\log(-\cC )\cdot \xz\rk+
\lk
f, \frac{i}{2}\rho(\tzx)\log(-\cC )\cdot i\frac{x^2}{2}\xz\rk.
\end{align*}
Then $F$ is analytic on $\HH\setminus\{i\alpha \mid \alpha\in[1,\infty)\}$ 
and the analytic continuation of the analytic function $G$ on $D\cap \HH$. \qed

In a similar manner 
we can see that $\rho(x^2)x\xz\in \rD(\tilde S_1)$ for $z\in \DD \cap \HH$, 
and 
\begin{align}
\label{gg}G_1(z)=\left(f, \tilde S_1 \rho(x^2)x\xz\right)=\left(f, \frac{i}{2}
\lk \rho(\tzx)\log(-\cC)\rk x\xz\right),\quad z\in \DD \cap \HH.\end{align}

\begin{lemma}\label{a2}
Let $f\in\LR$. Define 
$F_1(z)=(f, \tilde S_1 \rho(x^2)x\xz)$ for $z\in \HH\setminus\{i\}$. 
Then 
$F_1$ is analytic on $\HH\setminus\{i\alpha \mid \alpha\in[1,\infty)\}$. 
In particular 
$F_1$ is the analytic continuation of $G_1$ given by \eqref{gg} 
to $\HH\setminus\{i\alpha \mid \alpha\in[1,\infty)\}$. 
\end{lemma}
We define sesqui-linear forms 
$\tilde \cT_0$ and $\tilde \cT_1$ by
\begin{align*}
&\tilde \cT_0[\psi,\phi]=\half\left\{
\left(\tilde S_0\psi,\phi\right)+\left(\psi,\tilde S_0\phi\right)\right\},\quad \psi,\phi\in\kN_0,\\ 
&\tilde \cT_1[\psi,\phi]=\half\left\{
\left(\tilde S_1 \psi,\phi\right)+\left(\psi,\tilde S_1\phi\right)\right\},\quad \psi,\phi\in\kN_1. 
\end{align*}
The sesqui-linear form $\tilde \cT$ is defined by 
$$\tilde \cT=\tilde \cT_0\oplus \tilde\cT_1$$ 
with the domain $\kN\times \kN$. 
\begin{theorem}
The sesqui-linear form $\tilde \cT$ is an ultra-weak time operator of $h$ 
with the symmetric domain $\kN$ and 
the ultra-weak CCR-domain 
$\kN$. 
\end{theorem}
\proof
The theorem follows from Lemma \ref{a1}. 
\qed

Now we consider the matrix representation of $\tilde\cT$. 
We prove that $\cT[f_a,f_b]=\cT [x^a\xz,x^b\xi_z]$ diverges at $z=i$ in Theorem \ref{domain}. 
The next theorem is however to guarantee the existence of 
the analytic continuation of the map 
$z\mapsto \cT[f_a,f_b]$. 

\begin{theorem}\label{anacon}
Let $z\in \HH\setminus\{i\}$, $a,b\in\mathbb N$, $f_a=x^a\xz$ and $f_b=x^b\xz$. 
\begin{enumerate}[{\rm(1)}]
\item The following relation holds:
\begin{align*}
\tilde \cT [f_a, f_b]=
\left\{
\begin{array}{ll}
\displaystyle 
-\frac{i}{4} 
\lk \xz , \lkk \left(\tzx ^n x^{2m}-x^{2n}\tzx ^m\right) \log(-\cC )\rkk \xz \rk
& a=2n\ \mbox{and}\ b=2m,\vspace{5pt}\\
\displaystyle 
-\frac{i}{4} 
\lk \xz , \lkk \left(\tzx ^n x^{2m+2}-x^{2n+2}\tzx ^m\right) \log(-\cC ) \rkk \xz \rk
& \displaystyle \!\!\!{{a=2n+1\mbox{ and}}\atop{\ b=2m+1},\phantom{and}}\vspace{5pt}\\
\displaystyle 
0& \text{otherwise}.
\end{array}
\right.
\end{align*}
\item Suppose that $z=i\alpha $ with $\alpha\in (0,1)$. Then 
$\tilde \cT[f_a,f_b]= \cT[f_a,f_b]$. 
\item The map $z\mapsto \tilde \cT [f_a,f_b]$ is analytic on 
$\HH\setminus\{i\alpha \mid \alpha\in[1,\infty)\}$. 
In particular, $\tilde \cT [f_a,f_b]$ is the analytic continuation of $\cT [f_a,f_b]$. 
\end{enumerate}
\end{theorem}
\proof
(1) can be shown from the definition of $\tilde \cT$. 
(2) can be derived from $\tilde S_\#\rho(x^2)\xz=S_\# \rho(x^2)\xz$ for $z\in i(0,1)$. 
Finally (3) follows from Lemmas \ref{b2} and \ref{a2}. \qed

\section{Continuum limits}
In this section we consider the continuum limit of 
$\cT _\eps$ as $\eps\to0$. 
$T_\eps$ is formally written as 
$T_\eps=
-\half \frac{1}{\sqrt\eps}(\arctan\sqrt\eps t_0 +\arctan\sqrt\eps t_1)$. 
Since 
\[\lim_{\eps\to0}\half \frac{1}{\sqrt\eps}(\arctan\sqrt\eps x+
\arctan\sqrt\eps y)=\half (x+y)\quad (x,y\in\RR),\]
it can be expected that 
$\lim_{\eps\to0}T_\eps=\half(t_0+t_1)$ in some sense. 
Since subspaces $\kM_{\eps, \#}$ depend on $\eps$, 
to consider the continuum limit of $\cT _\eps$ we introduce 
$\kM_\#$ by $\kM_{\#,\eps}$ with $\eps$ replaced by $1$:
\begin{align*}
\kM_0&=\textrm{LH}\left\{x^{2n}e^{-\alpha x^2/2}\ \Big|\ n\in\mathbb N,\ \alpha\in(0,1)\right\},\\
\kM_1&=\textrm{LH}\left\{x^{2n+1}e^{-\alpha x^2/2}\ \Big|\ n\in\mathbb N,\ \alpha\in(0,1)\right\}.
\end{align*}
Let $\kM=\kM_0\oplus\kM_1$ and $\kD=\rD(S)\oplus \rD(S_1)$. 
Then the sesqui-linear form $\cT _\eps$ is also an ultra-weak time operator with 
the symmetric domain $\kD$ and 
the ultra-weak CCR domain 
$\kM$. We also extend the Aharonov-Bohm operator $T_{\rm AB}=\half(t_0+t_1)$ 
to an ultra-weak time operator. 
Let 
\begin{align*}
\cT_{0,\mathrm{AB}}[\psi,\phi]&=-\half \left\{(\psi,t_0\phi)+(t_0\psi,\phi)\right\},\quad
\psi,\phi\in \kM_0,\\
\cT_{1,\mathrm{AB}}[\psi,\phi]&=-\half \left\{(\psi,t_1\phi)+(t_1\psi,\phi)\right\},\quad 
\psi,\phi\in \kM_1.
\end{align*}
Define 
$\cT_{\mathrm{AB}}$ by 
$$
\cT_{\mathrm{AB}}=
\cT_{0,\mathrm{AB}}\oplus \cT _{1,\mathrm{AB}}.$$
The sesqui-linear form $\cT _{\mathrm{AB}}$ is an ultra-weak time operator 
of
$\half q^2$ with 
the symmetric domain $\kM$ and 
the ultra-weak CCR-domain $\kM$. 
\begin{theorem}
\label{cc}
For any $\phi,\psi\in \kM$, 
\begin{align*}
\lim_{\eps\to0}
\cT _\eps[\psi,\phi]=
\cT _{\mathrm{AB}}[\psi,\phi].
\end{align*}
\end{theorem}
\proof
Let $c_1,c_2,d_1,d_2\in\mathbb C$, 
$f_1=x^{2k}e^{-\alpha x^2/2}$, 
$f_2=x^{2l}e^{-\beta x^2/2}$, 
$g_1=x^{2m+1}e^{-\gamma x^2/2}$ and 
$g_2=x^{2n+1}e^{-\delta x^2/2}$. 
Set 
$\phi=c_1f_1+d_1g_1$ and $\psi=c_2f_2+d_2g_2$. 
It is sufficient to show that $\lim_{\eps\to0}\cT _\eps[\phi,\psi]=\cT _{\mathrm{AB}}[\phi, \psi]$.
From Theorem \ref{fab} it can be seen that 
\begin{equation*}
\cT_{0,\eps}[f_1, f_2]
=
\frac{i}{4\sqrt\eps}\left(\xi_{i\alpha ,1}, \left\{x^{2l}\taxo^k\log\frac{1+\sqrt\eps \alpha}{1-\sqrt\eps\alpha}-x^{2k}\tbxo^l\log\frac{1+\sqrt\eps \beta}{1-\sqrt\eps\beta}\right\}\xi_{\beta i,1}\right).
\end{equation*}
Expanding $\taxo^k\log
\frac{1+\sqrt\eps \alpha}{1-\sqrt\eps\alpha}$ on $x$, we have
\begin{align*}
 x^{2l}\taxo^k\log\frac{1+\sqrt\eps \alpha}{1-\sqrt\eps\alpha}&=x^{2(l+k)}\log\frac{1+\sqrt\eps \alpha}{1-\sqrt\eps\alpha}-k\sqrt\eps x^{2(l+k-1)}\frac{4}{1-\eps\alpha^2}+R^{}_{k,l,\alpha,\eps}(x),
\end{align*}
where $R^{}_{k,l,\alpha,\eps}(x)$ is a real valued function on $\mathbb R$ such that
$\lim_{\eps\to 0}\frac{1}{\sqrt\eps}R^{}_{k,l,\alpha,\eps}(x)=0$ 
for all $x\in\mathbb R$. 
Since $\frac{1}{\sqrt\eps} 
\log\frac{1+\sqrt\eps X}{1-\sqrt\eps X} 
\to 2X$ and $(1-\eps X^2)^{-1}\to 1$ as $\eps\to0$, 
we see that 
\begin{align*}
 \lim_{\eps\to0}\cT_{0,\eps}[f_1, f_2]&=\frac{i}{2}\left(\xi_{i\alpha ,1},x^{2(l+k-1)}\{\alpha x^2-2k-(\beta x^2-2l)\}\xi_{\beta i,1}\right)\\
 &=\frac{i}{2}\left(\left((\alpha x^2-2k)x^{2(k-1)}\xi_{i\alpha ,1}, x^{2l}\xi_{\beta i,1}\right)-\left(x^{2k}\xi_{i\alpha ,1},(\beta x^2-2l)x^{2(l-1)}\xi_{\beta i,1}\right)\right)\\
 &=-\frac{1}{2}\bigl((tf_1, f_2)+(f_1, tf_2)\bigr)\\
 &=\cT_{0,\mathrm{AB}}[f_1, f_2].
\end{align*}
In the same way as above, we obtain $\lim_{\eps\to0}\cT_{1,\eps}[g_1,g_2]=\cT_{1,\mathrm{AB}}[g_1,g_2]$.
Hence, for any $\psi,\phi\in \kM$, we conclude that 
$
\lim_{\eps\to0}
\cT _\eps[\phi,\psi]=
\cT _{\mathrm{AB}}[\phi, \psi]$.
\qed

\section{Time operators by POVM}
\label{povm2}
Let $\cH$ be a Hilbert space and $(\Omega, \cB_\Omega)$ a measurable space. 
An operator-valued set function $P$ on $\cB_\Omega$ is a POVM if and only if 
for every $A\in \cB_\Omega$, $P(A)$ is a bounded non-negative self-adjoint operator, and 
$$\cB_\Omega\ni A\mapsto \frac{(f, P(A)f)}{\|f\|^2}\in[0,1]$$ is a probability measure for any non-zero $f\in\cH$.
Note that $P$ is not necessarily projection-valued.

In this section we compare $T=T_{\eps=1}$ to the time operator 
$\TG$ derived through the POVM associated with $h=\half(p^2+q^2)$. 
The normalized eigenvector of $h$ associated with the eigenvalue 
$n+\half$ is denoted by $e_n$. 
Fot $t\in\RR$ and $N\in\NN$ we define 
$$\varphi_N(t)=\sum_{n=0}^N e^{-ith}e_n$$
and we set 
$$P_N(t)=\frac{1}{2\pi} (\varphi_N(t), \cdot) \varphi_N(t).$$
Let $\cB$ be the Borel $\sigma$-field on $[0,2\pi]$. 
The operator-valued set function $P_N$ on 
$\cB$ is defined by 
$$\cB\ni A\mapsto (f, P_N (A)g) = \int _{[0,2\pi]}\one_A(t) 
(f, P_N(t)g) dt.$$
Let $A,A'\in \cB$ and $A\subset A'$. Then 
$0\leq (f, P_N (A)f)\leq
(f, P_N (A')f)$. 

\begin{lemma}
For any $A\in \cB$ and any $f,g\in\cH$, 
$\lim_{N\to\infty} (f, P_N(A)g)$ exists.
\end{lemma}
\proof
It is immediate to see that 
\begin{align}
\label{bound}
(f, P_N (A)g) = 
\frac{1}{2\pi} \int _{A}\sum_{n=0}^N
(f,e_n)(e_n,g) dt+
\frac{1}{2\pi} \int _{A}\sum_{n\neq m}^N
e^{-it(n-m)}(f,e_n)(e_m,g) dt.
\end{align}
Suppose that $f=g$ and 
$A=[0,2\pi]$. Then 
$(f, P_N ([0,2\pi])f)=(2\pi)^{-1} \int _0^{2\pi}\sum_{n=0}^N
|(f,e_n)|^2 dt\leq \|f\|^2$. 
Using the polarization identity we can also see that for any $A\in \cB$, 
$\{(f, P_N(A)g)\}_N$ is a Cauchy sequence in $\CC$. Then the lemma follows. 
\qed

For each $A\in\cB$, by the Riesz representation theorem there 
exists a unique bounded non-negative self-adjoint operator $P(A)$ such that 
\begin{align}\label{povm}
(f, P(A)g)=\lim_{N\to\infty} (f, P_N(A)g).
\end{align}
Since $(f, P([0,2\pi])f)=\|f\|^2$, 
$P(\cdot)$ is a POVM on $([0,2\pi],\cB)$. 
Now the sesqui-linear form $\cT_G[f, g]$ is defined by 
\begin{align}\label{tg}
\cT_G[f, g]=\lim_{N\to\infty} \int_{[0,2\pi]}t\bigl(f, P_N(t) g\bigr) dt
\end{align}
with the domain $\LR\times\LR$. 
\begin{lemma}
The sesqui-linear form $\cT_G$ is bounded; 
$|\cT_G[f, g]|\leq 2\pi\|f\|\|g\|$ for $f,g\in\LR$. 
\end{lemma}
\proof
By a similar computation to \eqref{bound}, 
we obtain that $$
\cT_G[f, g]=\pi(f,g)+i
\sum_{n=0}^\infty\left(\sum_{m\neq n}\frac{(e_m,g) }{n-m}(f, e_n)\right)
$$
and 
$
|\cT_G[f, g]|\leq 
(2\pi) \|f\| \|g\|$. 
Here we used the inequality \cite[Theorem 294]{HLP34}:
$$\left|
\sum_{n=0}^\infty
\left(\sum_{m\neq n}\frac{x_n x_m}{n-m}\right)\right|\leq 
\pi 
\lk \sum_{n=0}^\infty x_n^2\rk ^{\han} 
\lk \sum_{n=0}^\infty y_n^2\rk ^{\han}.$$ 
Then the lemma is proven. 
\qed

The Riesz representation theorem yields again that 
there exists a unique bounded self-adjoint operator $\TG $ 
such that
 $\|\TG \|\leq 2\pi $ and 
\begin{align}\label{p1}
\cT_G[f, g]=(f, T_Gg). 
\end{align}
Formally writing $P(A)=\int_AdP$, we can express $\TG$ as 
\begin{align}\label{p2}
\TG=\int_{[0,2\pi]}tdP.
\end{align}
We shall show that $\TG$ is a time operator of $h$. 
\begin{lemma}
\label{63}
Let $f,g\in \rD(h)$. Then 
we have 
 $$\left(f, \bigl[h, P_N(t)\bigr]g\right)=-\frac{1}{i}\frac{d}{dt} \left(f, P_N(t)g\right).$$
 \end{lemma}
 \proof
Since
 $he^{-ith}e_n=-\frac{1}{i}\frac{d}{dt} e^{-ith}e_n$ in the strong sense, 
we have
 \begin{align*}
 (f, [h, P_N(t)]g)
&=
\frac{-1}{2\pi} \sum_{1\leq n,m\leq N}
 \lkk
 \frac{1}{i}\frac{d}{dt}
(f, e^{-ith} e_m)\cdot (e^{-ith}e_n,g)+
(f, e^{-ith} e_m)\cdot \frac{1}{i}\frac{d}{dt}
(e^{-ith}e_n,g)
\rkk\\
&=
\frac{-1}{i}\frac{d}{dt} \left(f, P_N(t) g\right).
 \end{align*}
 Then the lemma follows. 
 \qed

We define the unbounded operator $P_0$ by 
the strong limit:
$$P_0=\text{s-}\!\!\!\!\lim_{N\to\infty} P_N(0)=\text{s-}\!\!\!\!\lim_{N\to\infty}
\frac{1}{2\pi}
\left(\sum_{m=0}^N e_m,\cdot\right)\sum_{n=0}^N e_n.$$

\begin{lemma}
\label{galapon}
We have  
$[h, \TG ]=-i(\one-2\pi P_0 )$ on $\rD(h)$. 
\end{lemma}
\proof
Let $f,g\in \rD(h)$. 
By Lemma \ref{63} we have
\begin{align*}
(f, [h, \TG ]g)&=
\lim_{N\to\infty}
\int_{[0,2\pi]}t \left(f, \bigl[h, P_N(t)\bigr]g\right) dt=
-\lim_{N\to\infty}
\int_{[0,2\pi]}t \frac{1}{i}\frac{d}{dt} \left(f, P_N(t) g\right) dt\\
&=
-\lim_{N\to\infty}
\lkk
\frac{1}{i}
\left[t (f, P_N(t) g)\right]_0^{2\pi}
-
\int_{[0,2\pi]} \frac{1}{i}\left(f, P_N(t) g\right) dt\rkk \\
&=
i \lim_{N\to\infty}\left(f, 2\pi P_N(0)g\right)
-i(f,g). 
 \end{align*}
Then the lemma follows. 
\qed
\begin{lemma}
The operator $\TG$ is a bounded self-adjoint time operator of $h$, i.e., $[h, \TG ]=-i\one$ on $\textrm{LH}\{e_n-e_m\mid n\neq m\}$.
\end{lemma}
\proof
Since $P_0(e_n-e_m)=0$, the lemma follows from Lemma \ref{galapon}.
\qed

\begin{remark}
By the definition of $\TG $ 
we can see that
\begin{equation*}
\TG =
i\sum_{n=0}^\infty\left(\sum_{m\neq n}\frac{(e_m,\cdot) }{n-m}e_n\right)+\pi\one.
\end{equation*} 
This can be extended to a general self-adjoint operator $H$ 
possessing purely discrete spectrum: $\s(H)=\{E_n\}_{n=0}^\infty$ and $Hf_n=E_nf_n$. 
The time operators of $H$ of the form 
$$T_H=i\sum_{n=0}^\infty\left(\sum_{m\neq n}\frac{(f_m,\cdot) }{E_n-E_m}f_n\right)
$$
are studied in 
e.g., \cite{AM08,gal02a,gal02b}. 
\end{remark}

We can also define the sesqui-linear form associated with $\TG $ by 
$$\cT _G[\phi,\psi]=\half\{ (\phi, \TG \psi)+(\TG \phi,\psi)\}.$$ 
The sesqui-linear form $\cT _G$ is also an ultra-weak time operator of $h$ 
with the symmetric domain $\LR\times\LR$ and the ultra-weak CCR domain $\LR\times\LR$. 
Note that $\cT _G$ is a bounded sesqui-linear form.
\begin{theorem}\label{galapon2}
The sesqui-linear form $\cT$ is not equal to $\cT _{G}$. 
\end{theorem}
\proof
The sesqui-linear form $\cT _{G}$ is bounded, but $\cT$ is unbounded. Then 
the theorem follows. \qed

\section{Conclusion}
In this paper we discuss two time operators. 
One is the angle operator of the form
$$T_\eps=-\half \frac{1}{\sqrt\eps}(\arctan(\sqrt \eps q^{-1}p)+\arctan(\sqrt \eps pq^{-1}))$$
and the other is the operator of the form
\begin{equation*}
\TG =
i\sum_{n=0}^\infty\left(\sum_{m\neq n}\frac{(e_m,\cdot) }{n-m}e_n\right)+\pi\one.
\end{equation*} 
We give the firm definition of $T_\eps$ as a sesqui-linear form
and its matrix representation. 
Furthermore the analytic continuation of the matrix representation 
is obtained. 
We also show that $T_\eps$ converges to 
the Aharonov-Bohm operator $T_{\rm AB}$ as $\eps\to0$ in the sense of sesqui-linear forms. 
Finally it is shown that $T_{\eps=1}\neq \TG$ in the sense of sesqui-linear forms. 
As far as we know these are new.

We introduced the phase operator $\hat\phi$ in 
\eqref{phase} as a time operator of the number operator $N=\add a $. 
$N$ and $\hat h=\half (p^2+q^2)$ are simply 
related as $N+\half\one =\hat h$. 
The rigorous definitions of both $\log a$ and $\log \add$ are however not established. 
In particular it is not obvious to define $\log \add$ as a densely defined operator in $\LR$. 
Let $\hat T$ be \eqref{hatt}. 
One can expect however in e.g., \cite{SG64} that 
$\hat\phi=\hat T+G(\hat h)$ with some function $G$. 
Needless to say it is far away from mathematics. 
We shall discuss this in the second paper \cite{HT22}. 
Furthermore
we shall also show in \cite{HT22} the existence of the interpolation 
connecting the angle operator $T_{\eps=1}$ with $\TG$. 
\vspace{1cm} \\
{\bf Acknowledgements:}
FH is financially supported by JSPS KAKENHI 20K20886 %
 and JSPS KAKENHI 20H01808. \\
{\bf AUTHOR DECLARATIONS}\\
{\bf Conflict of Interest}
The authors have no conflicts to disclose. \\
{\bf Author Contributions}
Fumio Hiroshima: 
Conceptualization (equal);
Investigation (equal); 
Writing--review \& editing (leading). 
Noriaki Teranishi: 
Conceptualization (equal);
Investigation (equal); 
Writing--review \& editing (supporting).\\
{\bf DATA AVAILABILITY}
Data sharing is not applicable to this article as no new data were created or analyzed in this study.

\bibliographystyle{plain}
{\small \bibliography{hiro7}}

\end{document}